\documentclass[11pt,preprint]{aastex}

\newcommand{\etal}{{\it et~al.}}

\usepackage{rotating}

\begin{document}

\title{Thermophysical modeling of NEOWISE observations of DESTINY$^+$
  targets Phaethon and 2005 UD.}

\author{Joseph R. Masiero\altaffilmark{1}, E.L. Wright\altaffilmark{2}, A.K. Mainzer\altaffilmark{1}}

\altaffiltext{1}{Jet Propulsion Laboratory/California Institute of Technology, 4800 Oak Grove Dr., MS 183-301, Pasadena, CA 91109, USA, {\it Joseph.Masiero@jpl.nasa.gov}}
\altaffiltext{2}{University of California, Los Angeles, CA, 90095, USA}

\begin{abstract}

Thermophysical models allow for improved constraints on the physical
and thermal surface properties of asteroids beyond what can be
inferred from more simple thermal modeling, provided a sufficient
number of observations is available.  We present thermophysical
modeling results of observations from the NEOWISE mission for two
near-Earth asteroids which are the targets of the DESTINY$^+$ flyby
mission: (3200) Phaethon and (155140) 2005 UD.  Our model assumes a
rotating, cratered, spherical surface, and employs a Monte Carlo
Markov Chain to explore the multi-dimensional parameter space of the
fit.  We find an effective spherical diameter for Phaethon of
$4.6^{+0.2}_{-0.3}$km, a geometric albedo of $p_V=0.16\pm0.02$, and a
thermal inertia $\Gamma=880$ $^{+580}_{-330}$, using five epochs of
NEOWISE observations.  The best model fit for (155140) 2005 UD was
less well constrained due to only having two NEOWISE observation
epochs, giving a diameter of $1.2\pm0.4~$km and a geometric albedo of
$p_V=0.14\pm0.09$.
  
\end{abstract}

\section{Introduction}

Near-Earth asteroids (NEAs) are compelling targets for study because
of the hazard they pose to Earth due to their proximity to Earth's
orbit.  NEAs are remnants from the formation of the Solar system,
having recently been injected into near-Earth space from more distant
reservoir populations \citep{binzel15}.  Their complex history is
reflected in the wide diversity of physical characteristics they are
observed to have: internal structures from rubble piles to monoliths
\citep{fujiwara06,polishook17}; shapes ranging from nearly spherical
to elongated bi-lobes \citep{busch11,shepard18}; thermal histories
from pristine to highly evolved \citep{barucci18,popescu18}.  The
natural link between NEAs and meteorites in worldwide collections
provides a unique (though biased by atmospheric entry) ground-truth
against which to compare NEA observations
\citep[e.g.][]{jenniskens09}, while the same proximity of NEAs that
makes them a hazard also means they are convenient targets for
spacecraft missions \citep[e.g.][]{near,hayabusa,hayabusa2,orex}.

The DESTINY$^+$ mission, currently being developed by JAXA, will
conduct a flyby of NEA (3200) Phaethon and (155140) 2005
UD\footnote{\it https://destiny.isas.jaxa.jp/science/}
\citep{arai18,arai19}.  Phaethon has been linked to the Geminid meteor
shower \citep{gustafson89}.  There has been no evidence of
sublimation-based cometary activity, however the brightening of this
object at perihelion \citep{jewitt10,li13} has been linked to thermal
cracking and dust lofting \citep[e.g.][]{delbo14} though the observed
activity is not sufficient to explain the mass of the meteor shower.
Previous observations have shown that Phaethon is a B-class asteroid
based on visible light spectroscopy \citep{bus02,neesePDS}, with a
moderate geometric visible albedo of $\sim0.11$ \citep{tedesco04}.

The NEA (155140) 2005 UD has been linked to Phaethon both dynamically
and spectroscopically \citep{ohtsuka06,jewitt06}, implying that these
two objects potentially formed from a single precursor object.  Given
the similarity in their orbits, it is possible that these two
objects separated after moving into near-Earth orbital space. 
Observations from DESTINY$^+$ would provide critical
information for understanding the origin of these bodies.

Here we present new results for both of these objects, providing
constraints on their physical properties based on the thermal emission
observations from the NEOWISE survey data.

\section{Data}

We base this study on observations from the Wide-field Infrared Survey
Explorer (WISE) satellite \citep{wright10} which was used to conduct
the Near-Earth Object WISE (NEOWISE) surveys
\citep{mainzer11,mainzer14} during the cryogenic, 3-band cryogenic,
post-cryogenic \citep{cutri12}, and Reactivation mission phases
\citep{cutri15}.  WISE was launched on 14 Dec 2009, and surveyed the sky
from a Sun-synchronous polar orbit starting 7 Jan 2010, using
beamsplitters to simultaneously take images at four infrared
wavelengths: $3.4~\mu$m, $4.6~\mu$m, $12~\mu$m, and $22~\mu$m (known
as W1, W2, W3, W4, respectively).  After the outer cryogen tank was
exhausted on 6 Aug 2010 the optics warmed up and the W4 channel was no
longer able to be used to obtain survey data.  W3 continued operation
as the background temperature rose with decreasing exposure times
until 29 Sept 2010 when the inner cryogen reserve was exhausted,
causing the detectors to warm up.  Operations were continued in 2-band
post-cryo operations until 1 Feb 2011 when the telescope was put in
hibernation.  The 2 band survey was resumed 13 Dec 2013, and is
currently on-going having recently completed 5 years of restarted
operations.

This long operation period in nearly unchanging environmental
conditions has resulted in some of the asteroids and comets observed
having multiple detections at a range of orbital locations and viewing
geometries, producing a well-calibrated and uniform data set.  Using
the WISE Catalog Search Tool at the Infrared Science Archive ({\it
  https://irsa.ipac.caltech.edu}) we performed a moving object search
for all detections of our two objects of interest over all mission
phases: (3200) Phaethon and (155140) 2005 UD.  We find five epochs
with detections of Phaethon in at least one band at high SNR ($>5$),
one from the very beginning of the cryogenic survey and four from the
Reactivation mission.  For 2005 UD, we find two observation epochs,
both during the Reactivation mission.  We present the details of these
data in Table~\ref{tab.data}.  For detections from the Reactivation
data, we use the reported AllWISE atlas source search results to
filter out detections of background static astrophysical sources,
removing any detection with an atlas counterpart brighter than one
magnitude below the brightest band of the nominal moving object
detection (W2 for objects near the Earth's orbit).

Following the recommendations of the WISE Explanatory Supplement
\citep{cutri12}, we set a minimum error of $0.03$ mag for all
detections.  For the most recent Phaethon observing epoch, the
measured W2 magnitude was in the saturated regime.  As discussed in
\citet{cutri12}, Section 6.3.c.i.4, and related Figure 6.3.8b, W2 band
measurements for stars with magnitudes brighter than $\sim6~$mag begin
to show a linear deviation in their $K_s-$W2 color, resulting in a
flux over-estimation for saturated sources.  This is potentially due
to a flux-dependent PSF shape in W2, but no cause for this effect is
known.

We attempt to correct this flux overestimation by making a linear fit
to the $K_s-$W2 color vs W2 magnitude plot in the explanatory
supplement using the following procedure.  For detections with
$W2<6.1~$mag:

\begin{equation}
  W2_{out} = 0.788~W2_{in} + 1.29
  \label{eq.w2corr}
  \end{equation}

Our observations of these two targets span a range of phase angles
from $25^\circ - 77^\circ$.  In \citet{wright07}, the thermophysical
model (TPM) we use here was employed for data up to phase angles of
$130^\circ$, and thus the relatively large phase angles used here
should not be beyond the bounds of the model. \citet{mommert18} showed
through theoretical analysis that thermal model accuracy can be lower
at extremely high phase angles, so care must be taken when
interpreting these cases.

\begin{sidewaystable}[ht]
  \begin{center}
    \scriptsize
\caption{NEOWISE observing epochs used for thermophysical modeling}
\vspace{1ex}
\noindent
\begin{tabular}{cccccccccccc}
\tableline
Target  &  MJD   &  RA & Dec &  R$_{helio}$  & $\Delta$ & phase angle & Number of  & $<W1>$  & $<W2>$  & $<W3>$  & $<W4>$ \\ 
        & midpoint & (deg) & (deg)  &   (AU)          &  (AU)      &  (deg)             & Detections & (mag)  & (mag)  & (mag)  & (mag) \\ 
\tableline
(3200) Phaethon  & 55203.257 &   9.794 & +22.587 & 2.3170 &  2.07544 & 25.1 & 3 & 17.19 $\pm$ 0.47$^\dagger$ & 18.13 $\pm$ 2.97$^\dagger$ & 8.77 $\pm$ 0.07 & 6.14 $\pm$ 0.07\\
(3200) Phaethon  & 57035.346 &  21.646 & +16.851 & 1.3274 &  0.83122 & 47.7 & 7 & 14.34 $\pm$ 0.04 & 11.26 $\pm$ 0.03 & ... & ...\\
(3200) Phaethon  & 57663.549 & 239.600 & +84.379 & 1.0813 &  0.40568 & 67.7 & 3 & 12.42 $\pm$ 0.03 &  9.18 $\pm$ 0.03 & ... & ...\\
(3200) Phaethon  & 57729.562 & 350.026 & +22.871 & 1.8323 &  1.33255 & 31.5 & 27 & 16.16 $\pm$ 0.09 & 14.01 $\pm$ 0.05 & ... & ...\\
(3200) Phaethon  & 58104.215 &   3.777 & +25.015 & 1.0067 &  0.06905 & 68.9 & 1 &  7.74 $\pm$ 0.03 &  3.53 $\pm$ 0.03$^{\dagger\dagger}$ & ... & ...\\

\tableline

(155140) 2005 UD & 57750.591 & 357.719 & +47.069 & 1.3626 &  0.78202 & 45.2 & 7 & 16.78 $\pm$ 0.35 & 13.96 $\pm$ 0.13 & ... & ... \\
(155140) 2005 UD & 58383.429 &  89.606 &  +5.946 & 1.0312 &  0.24469 & 76.7 & 3 & 13.94 $\pm$ 0.13 & 10.68 $\pm$ 0.06 & ... & ... \\

\tableline
\end{tabular}
This table describes the different epochs of observation used for our
modeling.  Each row gives the mean MJD, right ascension and
declination of observation, distance to the sun and spacecraft,
magnitude in each WISE band, and the number of times the object was
detected in at least one band for each epoch.  {$^\dagger$}This is an
upper limit extracted using forced photometry by the WISE photometric
pipeline.\\{$^{\dagger\dagger}$}Saturated detection corrected
following Equation~\ref{eq.w2corr}\\`...' indicates no measurement
available
\label{tab.data}
\end{center}
\end{sidewaystable}

\clearpage

\section{Thermophysical Model}

\subsection{Model Description}
In order to simultaneously fit all observations of these two NEAs, we
employ a thermophysical model to constrain the diameter, albedo,
thermal inertia, pole position, and surface cratering fraction
(analogous to surface roughness in other TPM implementations).  For
our TPM we employ a rotating, crater model described in
\citet{wright07} and \citet{koren15}, using a Monte Carlo Markov Chain
(MCMC) to more efficiently explore the multi-dimensional parameter
space.  Many previous analyses of NEOWISE data
\citep[e.g.][etc.]{mainzer11neo,masiero18} have relied on the
Near-Earth Asteroid Thermal Model \citep[NEATM;][]{harris98}, which
has fewer free parameters and runs orders of magnitude faster.
However, NEATM relies on a non-physical ``beaming parameter'' that
varies with phase and observing circumstance as well as the actual
thermal, optical, and mechanical properties of the object
\citep{wright18}.  It is difficult to use beaming to infer these
physical properties beyond identifying outliers that are candidate
metallic objects or rapid rotators \citep{harris14}.

Our TPM uses as inputs the mean flux at each observing epoch,
including forced photometry derived upper-limit fluxes
reported by the WISE and NEOWISE source catalogs, as well as the
midpoint modified Julian date, the observed RA and Dec, the
observer-to-asteroid distance (queried from the JPL
Horizons\footnote{\it https://ssd.jpl.nasa.gov/horizons.cgi} system
for each epoch), the visible absolute $H_V$ magnitude, and the
rotation period if available.  For these fits, we use rotation periods
of 3.603958 hours for Phaethon and 5.231 hours for 2005 UD
\citep{lcdb}.

To account for the color corrections necessary to properly model the
WISE passbands, we use the analytical function described in
\citet{wright13} to calculate the output magnitude from the model for
each band $N$:

\begin{equation}
  mag_N = -2.5 \log\left[  a_{N1} F_{\nu}(\lambda_{N1}) + a_{N2} F_{\nu}(\lambda_{N2}) \right]
\end{equation}

where $mag_N$ is the output color-corrected magnitude, $a_{N1}$ and
$a_{N2}$ are the weights, and $F_{\nu}(\lambda_{N1})$ and
$F_{\nu}(\lambda_{N2})$ are the model fluxes at the two characteristic
wavelengths evaluated.  We use weights:
  
\[a_{11} = 0.51167, a_{12}=0.47952\]
\[a_{21} = 0.3165, a_{22}=0.6778\]
\[a_{31} = 0.3775, a_{32}=0.5188\]
\[a_{41} = 0.4554, a_{42}=0.5351\]

and wavelengths:

\[\lambda_{11} = 3.0974, \lambda_{12}=3.6298\]
\[\lambda_{21} = 5.0450, \lambda_{22}=4.4130\]
\[\lambda_{31} = 14.6388, \lambda_{32}=10.0348\]
\[\lambda_{41} = 20.4156, \lambda_{42}=23.5869\]

The $H$ absolute visible magnitude we use to fit the
  visible-light component of the spectrum is drawn from the most
  recent value published, allowing us to constrain
the geometric visible albedo and the ratio between the visible and
infrared ($3.4~\mu$m) albedos.  In this work, we use
$H=14.31\pm0.03$ for Phaethon \citep{hanus16} and
$H=17.3\pm0.2$ for 2005 UD (from the Minor Planet Center\footnote{\it
  https://minorplanetcenter.net}).  Note that these input H values are
treated as observational measurements along with IR data during the
MCMC, so the $p_V$ and $D$ parameters in the model may not exactly
reproduce the input.

Our TPM assumes a variable fraction of the surface is covered by
craters, which provide self-heating to unlit regions.  These craters
are not meant to represent the actual fraction of the surface with
visible impact markings, but rather provide a similar effect to
``roughness'' parameters used in other TPM implementations
\citep[e.g.][and references therein]{delbo15}.

The MCMC uses the two-Rayleigh fit to the NEO albedo distribution from
\citet{wright16} as the prior on the albedo distribution.  The prior
for the $p_{IR}/p_V$ ratio is assumed to be log-normal.  Priors on
other distributions are assumed to be uniform, with the diameter and
thermal inertia $\Gamma$ priors uniform in $\log$ space, the crater
fraction uniform in linear space, and the pole distribution uniform
over $4\pi$ steradians.  Our routine performs 72,900 accepted steps in
the MCMC, and reports the posterior parameter distribution based on
the last 48,600 steps.  The median plus the $16^{th}$ and $84^{th}$
percentile for each parameter are reported, along with the best fit
model.  The best fit is determined by the $\chi^2$ at each step of the
MCMC, which is the sum of the squares of the infrared flux deviations
plus the deviation of the model $H$ magnitude from the input $H$, each
weighted by the associated measurement uncertainty.

\subsection{Model Validation}

In order to validate the outputs of our TPM, in particular the
diameter accuracy, we conducted model fits to 23 objects that were
observed at multiple epochs by NEOWISE and also had high quality size
measurements from independent methods.  In this case we selected 22
Main Belt asteroids that had high quality occultation measurements
(quality code $U=3$ or $U=4$) from \citet{occultation_pds}, as well as
the NEO (433) Eros, which has both high quality occultation
measurements as well as 3D shape model information from the {\it NEAR}
spacecraft rendezvous \citep{veverka00}.

Table~\ref{tab.occul} shows the objects used for this validation
test.  The standard diameter (D$_{std}$) used for comparison is the
error-weighted average of the circular-equivalent radius of the
ellipse fit to the occultation, for all U=3 and U=4 occultations.  In
the case of (433) Eros, we also show the spherical equivalent diameter
from the 3D ellipsoid from the {\it NEAR} observations.  We also list
the size reported by our TPM (D$_{TPM}$) and the number of epochs of
NEOWISE data used for the fit.  All objects were observed once or
twice during the 2010-2011 mission phase, with the rest of the epochs
coming from the reactivation mission.

For both asteroids (393) and (554), one observation epoch was obtained
during the end of the 3-band cryo mission, when the integration time
had been reduced to $1.1$ seconds due to the increasing background
\citep{cutri12}.  The W2-W3 color for these two objects in this epoch
were significantly different from what was expected based on other
Main Belt asteroids observed during the cryogenic mission.  This is
likely a symptom of the changing zero points and calibrations during
the end stage of warm up.  As the W3 zero points are not
well-determined for these cases, we have excluded them from our
modeling effort, as noted in the table. The remaining detections for
the Main Belt objects were typically in the saturated regime for the
cryogenic W3 and W4 detections, and in some cases in the W2 detections
as well.  Because of this, they are not ideal calibrators; however,
the vast majority of objects with high-quality size measurements from
occultations are these large MBAs.  Therefore, they represent a
reasonable validation set, but caution is recommended.

\begin{figure}[ht]
\begin{center}
\includegraphics[scale=0.6]{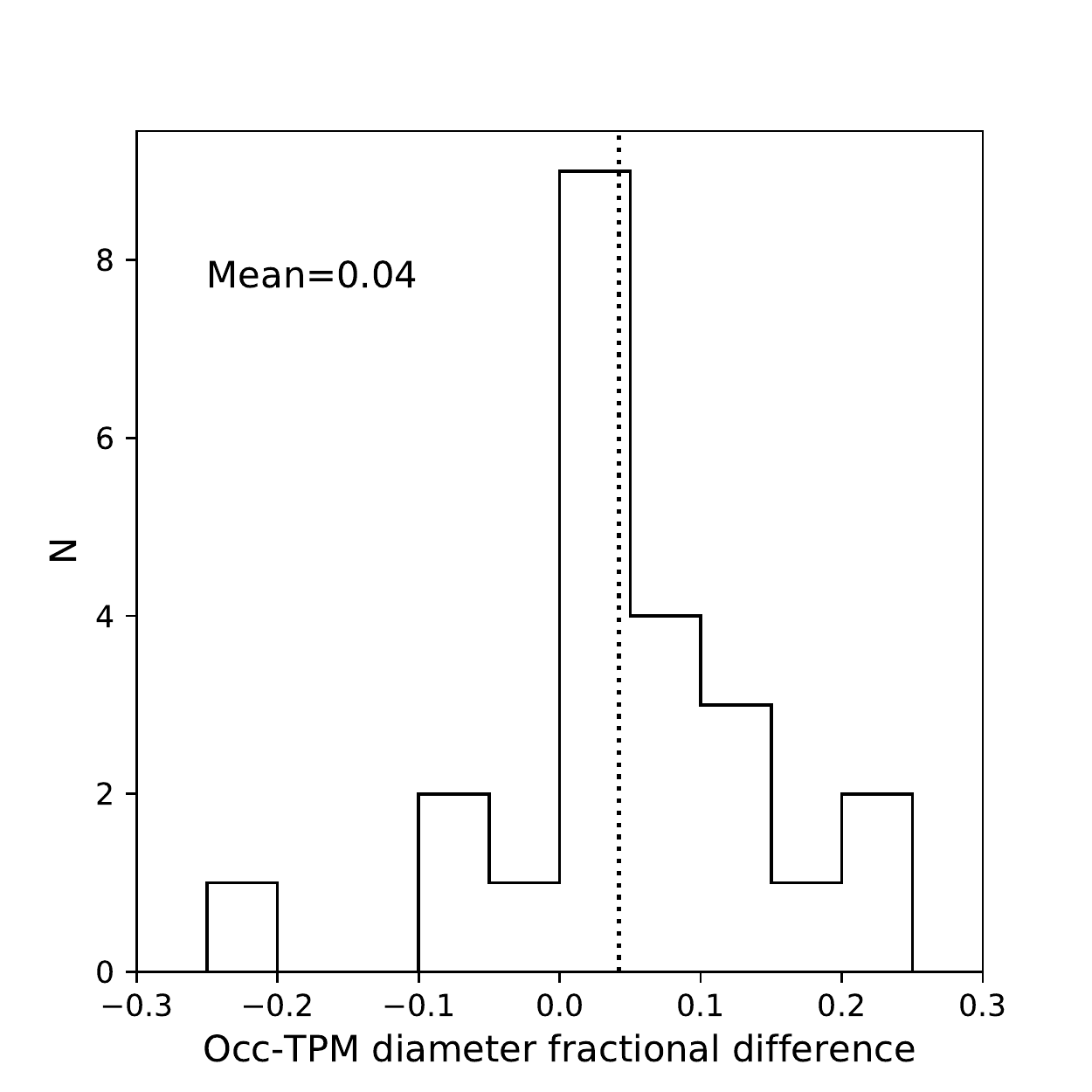}
\protect\caption{Fractional difference between occultation-derived
  sizes and TPM-derived sizes for 23 asteroids with sufficient data
  for both determinations.  The mean offset of $0.04$ is indicated by
  the dotted line.}
\label{fig.occul_hist}
\end{center}
\end{figure}

Additionally, as (433) Eros is a highly elongated object and the
occultation data appear to be from a viewing geometry oriented along
the long axis of the body, a more accurate comparison for the results
from the TPM are to the spherical equivalent size.  We list both
diameters in Table~\ref{tab.occul}, but only include the comparison
with the {\it NEAR} data in Figure~\ref{fig.occul_hist}.  This figure
shows the fractional difference between the ground-truth sizes and the
TPM-derived sizes, as well as the mean of these differences.  The mean
offset is $4\%$ (with the TPM results being slightly smaller), with
16$^{th}$ and 84$^{th}$ percentiles of $-6\%$ and $+10\%$.

It should be noted that while the measurement error on occultation
chords is typically smaller than $10\%$ (set by the speed of frame
acquisition and length of the occultation), inferred sizes are subject
to projection effects and can be significantly offset from the
spherical equivalent size for elongated objects.  The deviation in the
occultation size of Eros from the {\it in situ} measurements is an
example of this situation.  Thus, while this validation process shows
that for the bulk population the TPM-derived spherical equivalent
diameters agree with the averaged occultation diameters, significant
offsets are possible and we cannot say for individual cases where
there is a discrepancy which size determination is more reliable.

\begin{table}[ht]
\begin{center}
\caption{Thermophysical model fits for objects with ground-truth sizes}
\vspace{1ex}
\noindent
\begin{tabular}{cccc}
\tableline
Asteroid Number  &  D$_{std}$  &  D$_{TPM}$ & Number of NEOWISE epochs \\
                 &   km       &  km   &       \\
\tableline

17   &  70.8   & 77.0 $^{+4.3}_{-3.3}$ & 8  \\
51   &  146.1  & 138.4$^{+2.8}_{-3.3}$ & 8  \\
81   &  117.8  & 117.1$^{+2.0}_{-2.6}$ & 8  \\
95   &  140.0  & 139.7$^{+3.8}_{-4.7}$ & 9  \\
129  &  126.5  & 125.5$^{+2.3}_{-2.5}$ & 9  \\
134  &  116.6  & 107.8$^{+3.3}_{-5.7}$ & 8  \\
141  &  128.2  & 118.6$^{+5.6}_{-7.2}$ & 6  \\
144  &  141.4  & 122.1$^{+4.4}_{-3.5}$ & 7  \\
208  &  44.3   & 43.2 $^{+1.2}_{-1.1}$ & 8  \\
225  &  114.0  & 100.1$^{+5.3}_{-7.7}$ & 8  \\
238  &  145.3  & 148.3$^{+5.3}_{-25.2}$ & 9  \\
308  &  124.8  & 124.1$^{+2.7}_{-3.1}$ & 9  \\
393$^\dagger$  &  123.6  & 123.0$^{+8.0}_{-16.7}$ & 7 \\
404  &  100.5  & 97.2$^{+5.8}_{-19.3}$  & 9  \\
433  (occultation)&  10.1   & 17.5$^{+1.8}_{-1.5}$  & 5 \\
433  (NEAR) &  16.3   & 17.5$^{+1.8}_{-1.5}$  & 5 \\
526  &  44.6   & 42.0$^{+1.1}_{-1.2}$  & 9  \\
554$^\dagger$  & 103.4   & 84.5$^{+15.3}_{-19.9}$  & 7  \\
578  &  74.2   & 57.5$^{+1.7}_{-2.2}$  & 7  \\
580  &  49.5   & 47.8$^{+2.2}_{-3.7}$  & 8  \\
757  &  36.8   & 36.6$^{+1.0}_{-1.6}$  & 8  \\
874  &  43.9   & 54.6$^{+2.6}_{-1.9}$  & 8  \\
1263 &  44.2   & 33.3$^{+3.4}_{-4.7}$  & 6  \\
1366 &  29.8   & 26.8$^{+1.0}_{-1.0}$  & 7  \\

\tableline
\end{tabular}
\newline

{$^\dagger$}W3 measurement from end of 3-band cryo mission phase discarded as discussed in the text
\label{tab.occul}
\end{center}
\end{table}

\clearpage
  
\section{Results and Discussion}

\begin{figure}[ht]
\begin{center}
\includegraphics[scale=0.6]{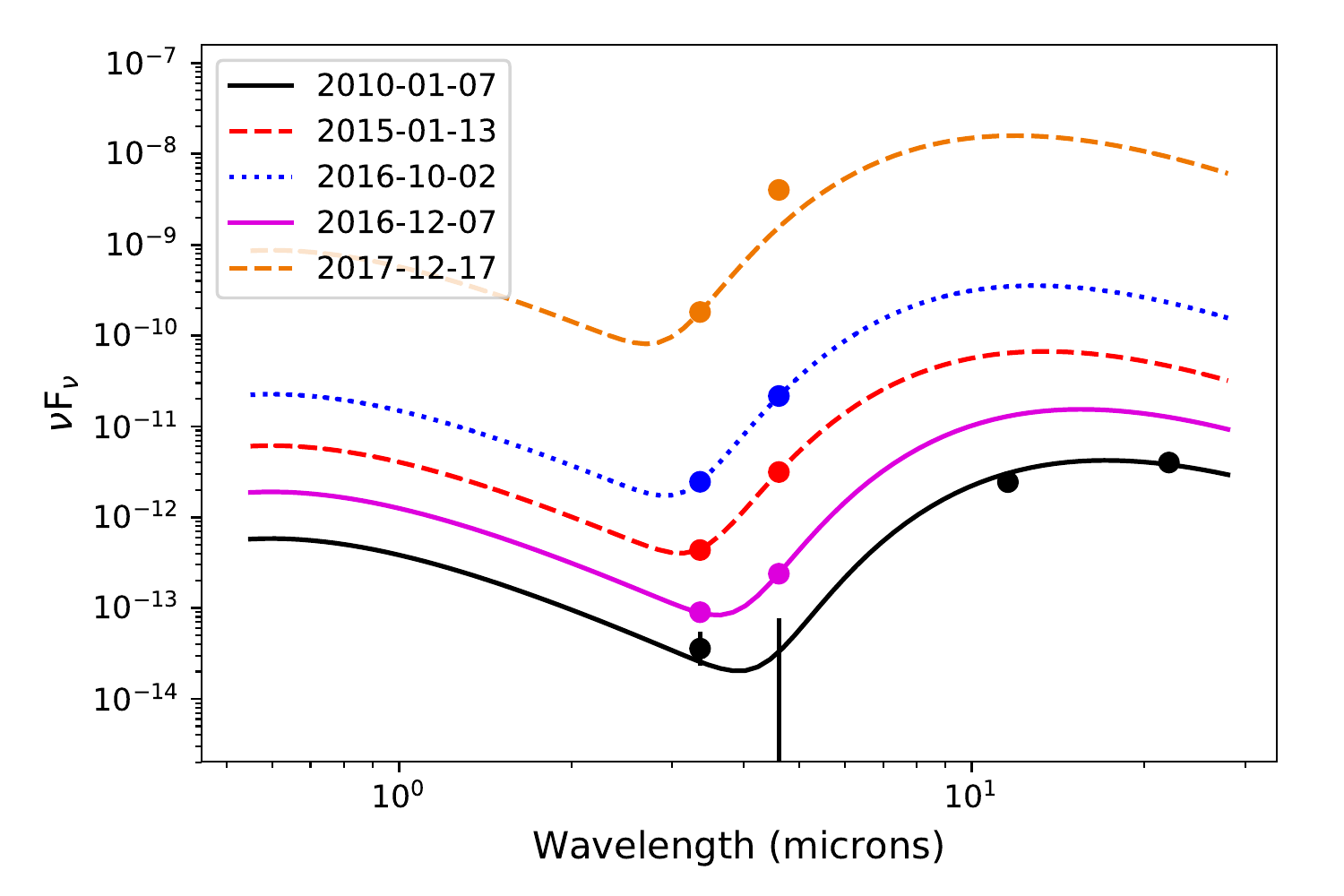}
\protect\caption{Predicted spectral energy distributions for the
  best-fit model of (3200) Phaethon at each observing epoch.  Lines
  show the model, points show the observations with associated error
  (typically the size of the point).}
\label{fig.phaethon_sed}
\end{center}
\end{figure}

\begin{figure}[ht]
  \begin{center}
    \includegraphics[scale=0.6]{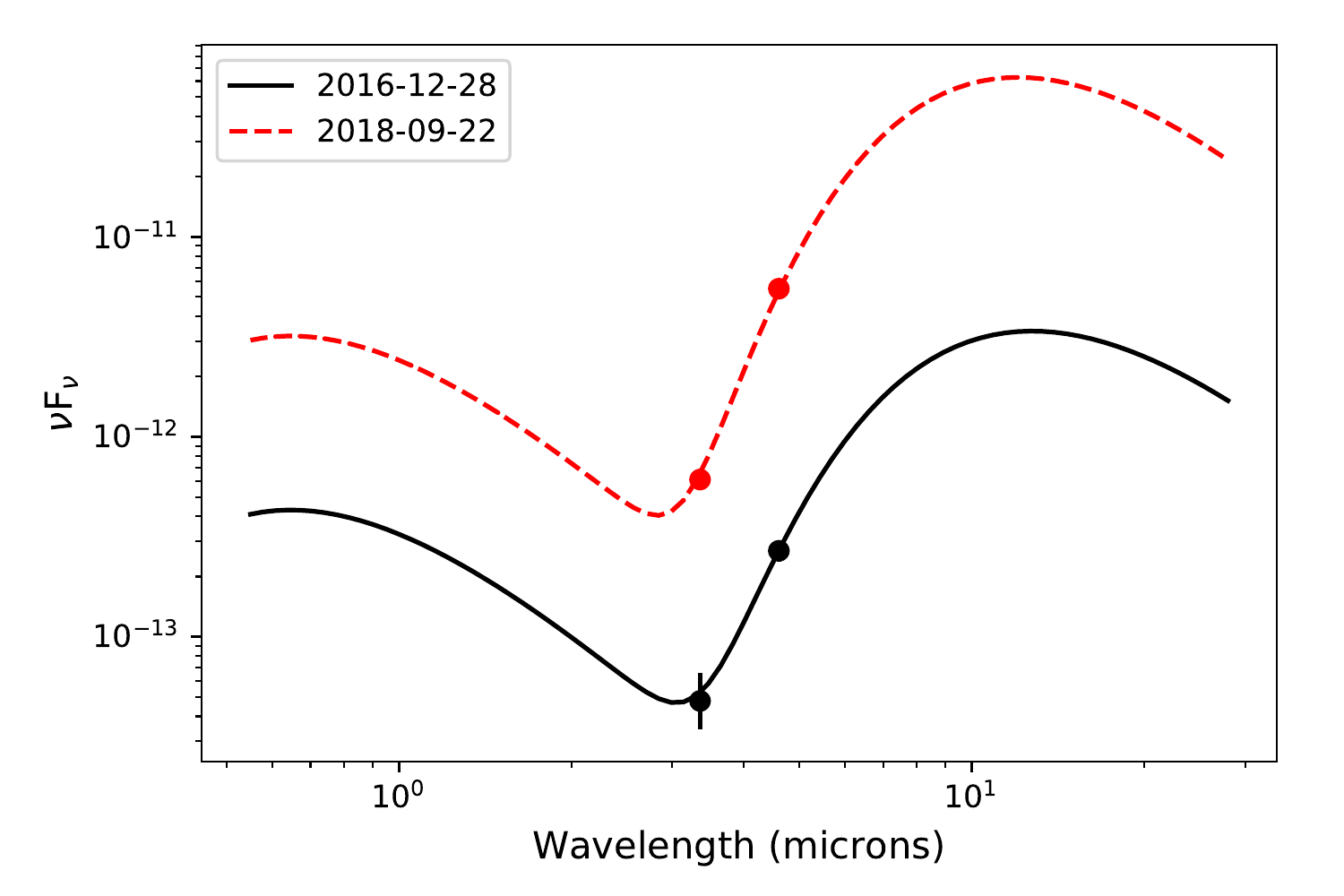}
\protect\caption{The same as Fig~\ref{fig.phaethon_sed} for (155140)
  2005 UD.}
\label{fig.2005UD_sed}
\end{center}
\end{figure}

Our best fit TPM for each asteroid is shown in
Figures~\ref{fig.phaethon_sed}~\&~\ref{fig.2005UD_sed}, and the fitted
model parameters are given in Table~\ref{tab.bestfit}.  Both objects
have moderate albedos, consistent with Pallas family-like B-type
objects \citep{ali-lagoa13}, and the ratios of their infrared albedo
to optical albedo are consistent with flat or weakly red spectral
slopes between $0.55~\mu$m and $3.4~\mu$m, also consistent with B-type
taxonomy \citep{mainzer11spectax}.  This slope follows the trend
across the surface of the reflectance spectrum change from blue to
more neutral with increasing wavelength \citep{kareta18}.

Previous TPM fits of Phaethon \citep{hanus16,hanus18} have resulted in
a slightly larger diameter than we find ($D=5.1\pm0.2~$km vs
$D=4.6^{+0.2}_{-0.3}~$km for our fit, a difference of
$\sim1.8~\sigma$), but comparable thermal inertias ($\Gamma=600\pm200$
J m$^{-2}$ s$^{-0.5}$ K$^{-1}$ vs $\Gamma=880^{+580}_{-330}$ J
m$^{-2}$ s$^{-0.5}$ K$^{-1}$ for our fit) despite using different data
sources (e.g. Spitzer and IRAS) and model assumptions.  We assume that
thermal inertia is constant for Phaethon, although as discussed in
\citep{delbo15} thermal inertia can vary as a function of temperature,
and thus heliocentric distance.  Because of the fixed WISE viewing
geometry, heliocentric distance is correlated with phase angle, which
plays an important role in constraining the cratering fraction of the
surface in our model.  Future work will investigate whether our
results would be improved by using a distance-corrected thermal
inertia for our model.

We note that our best fit model is unable to
reproduce the highly saturated flux in W2 for the 2017-12-17 observing
epoch (MJD$=58104.215$), even after our attempts to correct for the
non-linearity seen in W2 for very bright sources (see
Equation~\ref{eq.w2corr}).  Our best fit model falls $2.9~\sigma$
below the measured value.  As this epoch consists of only a single
observation, this result may be due to our model's assumption that the
object is spherical, or may indicate our bright source correction is
insufficient.

Recent radar data from Arecibo Observatory \citep{taylor19} show that
the observations are consistent with an sphere of diameter of
$6.2~$km, or a top-shaped figure with equivalent spherical diameter of
$5.5~$km, both of which are larger than either our result or the
results of \citet{hanus18}.  Further, \citet{taylor19} state that a
$5.1~$km sphere with a $3.6~$hr rotation period would have a maximum
Doppler broadening of the radar return significantly smaller than the
measured broadening. The radar data would thus imply that neither our
TPM fit nor the \citet{hanus18} TPM fit could be consistent with the
radar observations.  As both TPM fits are consistently below the radar
size measurement, this may point to a surface that behaves
thermophysically in a highly unusual way, or that Phaethon has a
non-spherical shape with a significant equatorial bulge \citep[see
  discussion in][]{taylor19}.

As NEOWISE only observed 2005 UD at two epochs, the TPM fit for this
object was not as well constrained as the fit for Phaethon.  We find
for 2005 UD a diameter of $1.2\pm0.4~$km and a geometric albedo of
$p_V=0.14 \pm 0.09$.  This is comparable to the diameter of $1.15 \pm
0.37$ found through a fit to the single epoch data from the fourth
year of NEOWISE data using NEATM (Masiero \etal in prep), and both
fits have similar uncertainties.

The posterior distribution of the rotational pole can be used to
constrain its position for objects with multiple observing geometries
that provide sufficient constraints. However, our observations of 2005
UD do not sufficiently constrain the pole, resulting in a significant
degeneracy in the on-sky pole position. Our model finds a mean
position for the pole of Phaethon at $350^\circ, -73^\circ$ (ecliptic
long/lat), but with significant extension of possible solutions to
$315^\circ, -45^\circ$ (near the best-fit solutions from
\citep{hanus18} and \citep{kim18}) as well as its antipode.  Thus our
position posterior does not provide an improved constraint over
previous work. The median of all MCMC solutions settled on a derived H
magnitude of $14.27$ for Phaethon and $17.35$ for 2005 UD, comparable
to the input values.

\begin{table}[ht]
\begin{center}
\caption{Best-fit thermophysical models from NEOWISE data}
\vspace{1ex}
\noindent
\begin{tabular}{cccccc}
\tableline
Target  &  Diameter  &  geometric visible albedo & $\Gamma$                      &  crater fraction  & $p_{IR}/p_V$ \\
        &   km  &                        & J m$^{-2}$ s$^{-0.5}$ K$^{-1}$  &                   &              \\
\tableline

(3200) Phaethon  & $4.6^{+0.2}_{-0.3}$ & $0.16\pm0.02$ & $880^{+580}_{-330}$ & $0.20^{+0.28}_{-0.14}$ & $0.8\pm0.06$ \\
(155140) 2005 UD & $1.2\pm0.4$ & $0.14\pm0.09$ &  ...                & $0.38^{+0.38}_{-0.28}$ & $1.41^{+0.36}_{-0.37}$ \\

\tableline
\end{tabular}
`...' indicates value was not constrained by the model
\label{tab.bestfit}
\end{center}
\end{table}

\begin{figure}[ht]
\begin{center}
\includegraphics[scale=0.6]{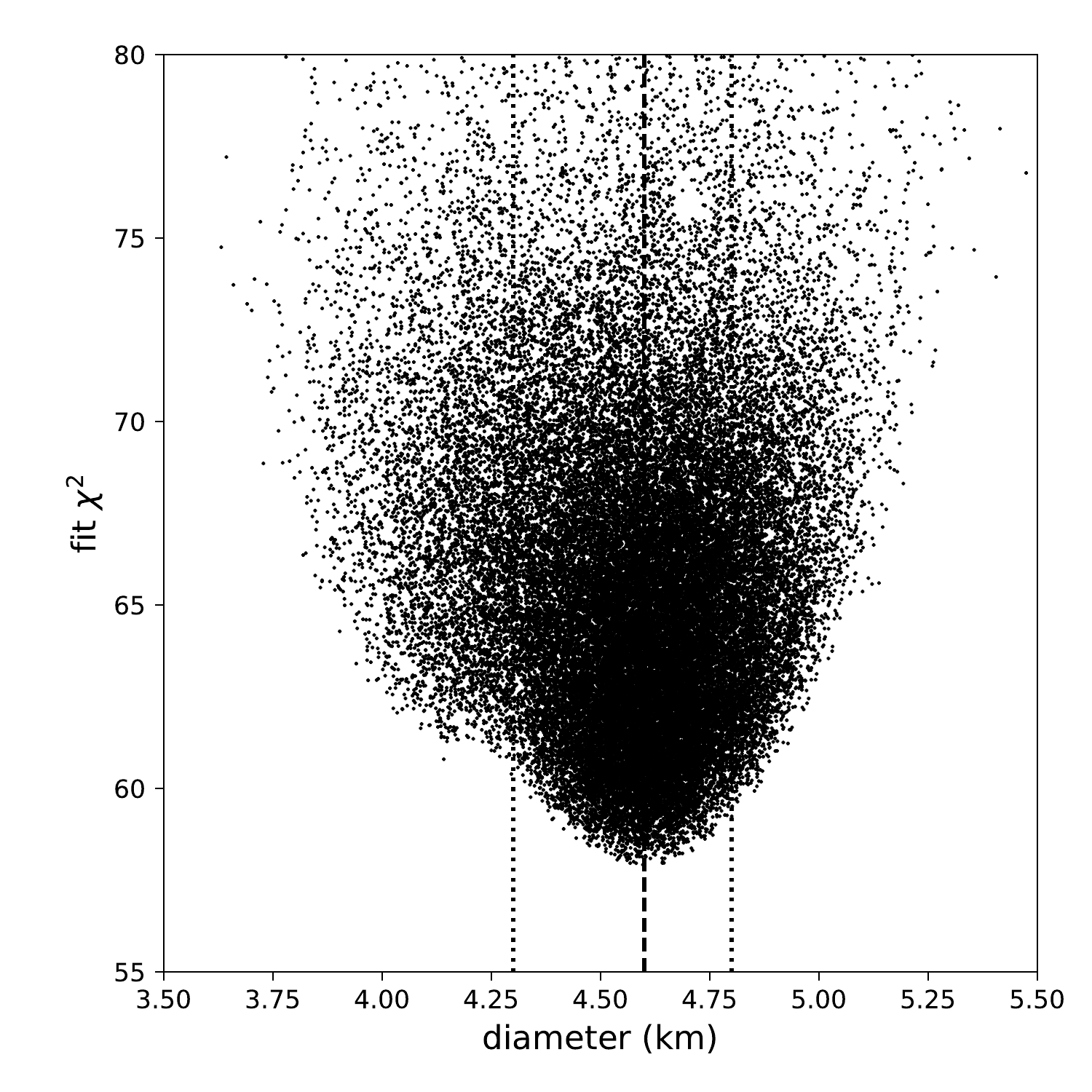} \protect\caption{Fitted
  diameter vs $\chi^2$ of all Monte Carlo trials for (3200) Phaethon
  (points). The dashed line shows the MCMC-weighted median of the
  fits, while the dotted lines show the $16^{th}$ and $84^{th}$
  percentile bounds.}
\label{fig.phaethon_dvchi}
\end{center}
\end{figure}

\begin{figure}[ht]
  \begin{center}
    \includegraphics[scale=0.6]{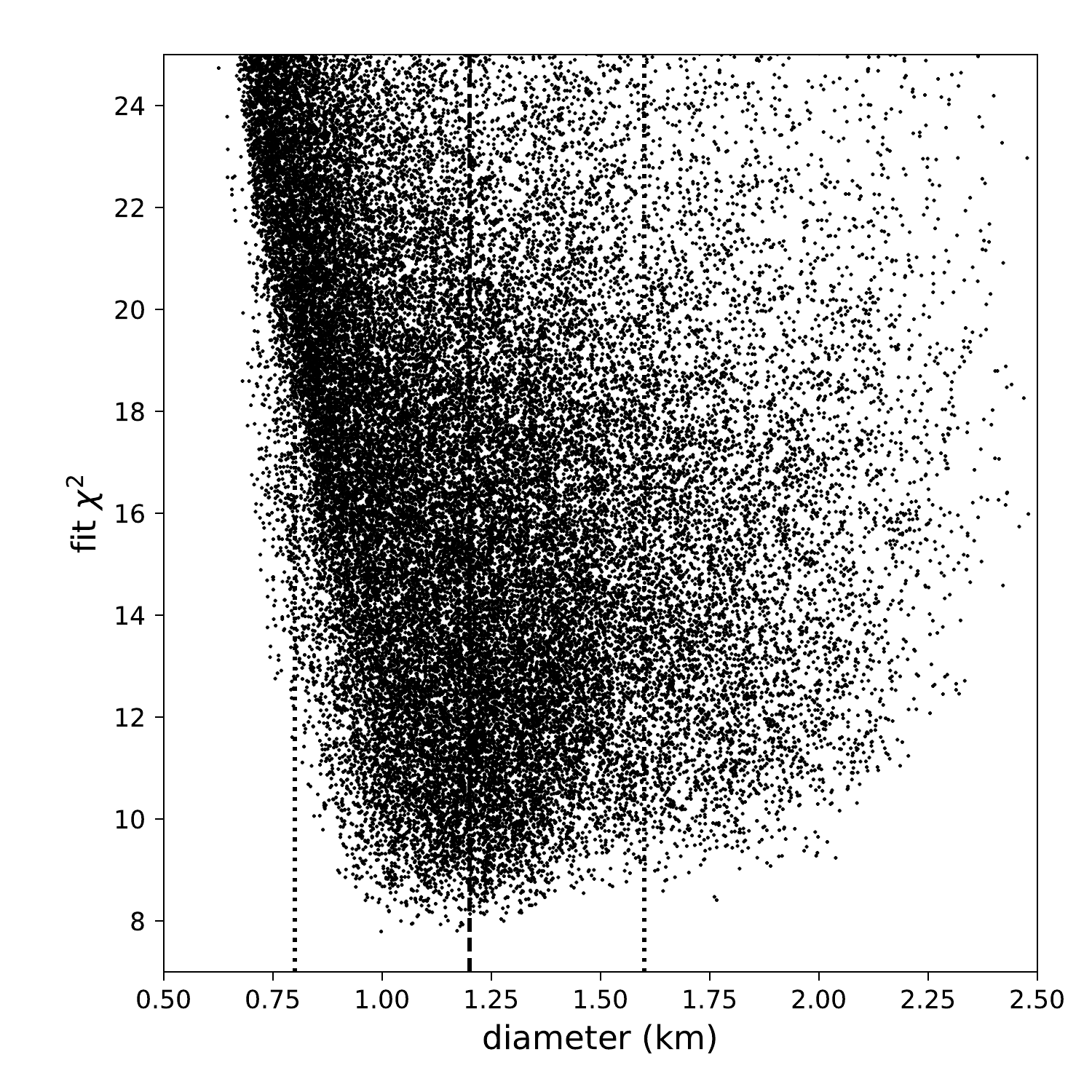}
    \protect\caption{The same as Fig~\ref{fig.phaethon_dvchi} for
      (155140) 2005 UD.}
\label{fig.2005UD_dvchi}
\end{center}
\end{figure}

We show in Figures~\ref{fig.phaethon_dvchi}~\&~\ref{fig.2005UD_dvchi}
a comparison of fitted diameter with the $\chi^2$ value of the fit for
each Monte Carlo iteration of the MCMC.  For Phaethon, the multiple
epochs enabled a better constraint on the diameter, resulting in a
tight clustering of Monte Carlo trial results.  For 2005 UD, the
diameter is less well-constrained due to having fewer observing epochs
(and thus measurements) available, resulting in a lower overall
$\chi^2$ (because of the fixed number of fitted parameters) but a
broader spread in values.  The large range of $\chi^2$ values at the
best fitting diameter is a result of variations in the other
parameters, such a thermal inertia and rotation pole.

\section{Conclusions}

Using a spherical, cratered MCMC thermophysical model, we have set
constraints on the physical properties for (3200) Phaethon and
(155140) 2005 UD based on thermal infrared measurements at multiple
observing epochs.  The model for Phaethon, based on one epoch of
4-band cryogenic WISE observations as well as four additional epochs
of 2-band NEOWISE reactivation observations, is well constrained and
offers improvements over simpler NEATM-based thermal model fits to the
same data.  We find that the median of all TPM fits gives an effective
spherical diameter of $4.6~$km $^{+0.2}_{-0.3}$, a geometric
  albedo of $p_V=0.16 \pm 0.02$, and a thermal inertia $\Gamma=880$
  $^{+580}_{-330}$, where the uncertainties quoted are for the
  $16^{th}$ and $84^{th}$ percentiles of the distributions, which are
  non-Gaussian.  These fits are consistent within $2 \sigma$ of
  previous TPM fits of other infrared data, but are $>3 \sigma$
  different from the maximum extent measured by radar.

The TPM for 2005 UD is less well-constrained as only two observing
epochs were available, both during the 2-band reactivation survey.
This median fit gives a size of $1.2\pm0.4~$km and a geometric albedo
of $p_V=0.14 \pm 0.09$, comparable to the quality of a NEATM-based
fit.  No useful constraint can be placed on the thermal inertia
$\Gamma$ from these data.  Two epochs of infrared observations
combined with a known rotation period is therefore likely to be the
bare minimum at which TPMs can produce results comparable to, or
better than, simple thermal models.  For observations with single
observation epochs, as is found for the majority of NEAs observed by
NEOWISE, thermal models like NEATM with fewer free parameters are
preferred due to the significantly smaller computational requirements,
while objects with many epochs covering multiple viewing geometries
will derive the most benefit from detailed thermophysical modeling.
In particular, thermophysical modeling will become a critical tool for
understanding NEAs as future infrared surveys such as the Near-Earth
Object Camera \citep[NEOCam;][]{mainzer15neocam} begin producing
multi-epoch infrared measurements of thousands of near-Earth
asteroids.

\section*{Acknowledgments}

The authors thank the anonymous referee for the detailed and helpful
comments that resulted in a significant improvement to this work.  JRM
thanks Shinsuke Abe for discussions that inspired this paper.  This
publication makes use of data products from the Wide-field Infrared
Survey Explorer, which is a joint project of the University of
California, Los Angeles, and the Jet Propulsion Laboratory/California
Institute of Technology, funded by the National Aeronautics and Space
Administration.  This publication also makes use of data products from
NEOWISE, which is a project of the Jet Propulsion
Laboratory/California Institute of Technology, funded by the Planetary
Science Division of the National Aeronautics and Space Administration.
This research has made use of data and services provided by the
International Astronomical Union's Minor Planet Center.  This
publication uses data obtained from the NASA Planetary Data System
(PDS).  This research has made use of the NASA/IPAC Infrared Science
Archive, which is operated by the Jet Propulsion Laboratory,
California Institute of Technology, under contract with the National
Aeronautics and Space Administration.

\end{document}